\documentclass[a4paper]{article}

\usepackage{amsfonts}
\usepackage{amssymb}
\usepackage{amsthm}
\usepackage{amsmath}
\usepackage{feynmp}
\usepackage{psfrag}
\usepackage{citesort}

\begin{document}

\title{\begin{flushright}
\small{ITP-UU-07/28}\\
\small{Spin 07/18}\\
\ \\
\ \\
\end{flushright}
Coordinate Bethe Ansatz for the String S-Matrix}
\author{M. de Leeuw%
    \thanks{M.deLeeuw@phys.uu.nl} \\
    \\
\textit{Institute for Theoretical Physics and Spinoza Institute,}\\ \textit{Utrecht University,}\\
\textit{3508 TD Utrecht, The Netherlands}}

\date{}

\maketitle

\begin{abstract}
We use the coordinate Bethe ansatz approach to derive the nested
Bethe equations corresponding to the recently found S-matrix for
strings in $AdS_{5}\times S^{5}$, compatible with centrally
extended $\mathfrak{su}(2|2)$ symmetry.
\end{abstract}

\section{Introduction}

Recently, there has been a lot of progress in understanding the
AdS/CFT correspondence. One of the most important developments was
the discovery of integrable structures on both sides of the
correspondence \cite{Bena:2003wd},\cite{Minahan:2002ve}.
Integrability provides new insights in how to calculate spectra
and how to study the correlation between the $AdS_{5}\times S^{5}$
string sigma model and its dual gauge theory. An important tool
used to solve quantum integrable systems is a technique called the
Bethe ansatz. The Bethe ansatz has been applied to a variety of
different problems and there are two main variations known; the
algebraic Bethe ansatz \cite{Faddeev:1996iy} and the coordinate
Bethe ansatz \cite{Yang:1967bm}.

\smallskip

On the gauge theory side of the AdS/CFT correspondence, integrable
structures emerged via spin chains \cite{Minahan:2002ve}; it was
observed that conformal operators of $\mathcal{N}=4$ SYM
correspond to eigenstates of an integrable spin chain at the
planar one-loop level. Furthermore, the scaling weights of the
conformal operators coincide with energy eigenvalues of the spin
chain Hamiltonian. There is much evidence that integrability on
the gauge theory side actually extends to all loop order and the
corresponding Bethe equations have been proposed for certain
asymptotic limits
\cite{Staudacher:2004tk}\nocite{Beisert:2004hm}-\cite{Beisert:2005fw}.

\smallskip

On the string theory side, integrability was exhibited for
classical strings on $AdS_{5}\times S^{5}$ \cite{Bena:2003wd}. One
important open question is whether integrability is inherited by
the quantum string. Assuming that this is the case for the full
quantum theory, a Bethe ansatz for the gauge-fixed string sigma
model was proposed \cite{Arutyunov:2004vx}. The construction of
\cite{Arutyunov:2004vx} is based on the knowledge of the
finite-gap solutions of the classical string sigma-model
\cite{Kazakov:2004qf}. The characteristic feature of the quantum
Bethe ansatz in comparison to the gauge theory Bethe ansatz is the
appearance of an additional scattering (dressing) phase
constructed as a two-form on the vector space of local conserved
charges. This dressing phase is universal and underlies the Bethe
equations of the full-fledged sigma model \cite{Beisert:2005fw}.
However, in contradistinction to the gauge theory side,
integrability at higher orders of string perturbation theory
remains conjectural. For recent advances in this direction based
on the direct world-sheet approach see
\cite{Hentschel:2007xn}\nocite{Klose:2006zd}-\cite{Roiban:2007jf}.

\smallskip

The S-matrix describing the scattering of world-sheet excitations,
respectively excitations of a certain spin chain, proved to be
crucial in determining the relevant spectrum in the large volume
(charge) limit \cite{Beisert:2005fw}-\cite{Beisert:2005tm}. This
S-matrix turns out to be severely restricted if one imposes
compatibility with the global symmetries of the model. It was
first shown for the $\mathcal{N}=4$ gauge theory that the relevant
(super)algebra was centrally extended
$\mathfrak{psu}(2|2)\oplus\mathfrak{psu}(2|2)$
\cite{Beisert:2005tm}. The same algebra also emerges for
superstrings on $AdS_{5}\times S^{5}$ as a symmetry algebra for
the light-cone gauge-fixed Hamiltonian \cite{Arutyunov:2006ak}. It
appears that both the two-particle S-matrix for superstrings on
$AdS_{5}\times S^{5}$ in the decompactifying limit and the
S-matrix for the $\mathfrak{su}(2|2)$ dynamic spin chain
\cite{Beisert:2005tm} can be uniquely determined up to an overall
phase factor by requiring invariance under this global symmetry
algebra.

\smallskip

By imposing the requirement of crossing symmetry, which is a
common property of relativistic field theories, one derives
constraints on the dressing phase \cite{Janik:2006dc}. Based on
earlier work
\cite{Arutyunov:2006iu}\nocite{Freyhult:2006vr}-\cite{Hernandez:2006tk},
an explicit all-order perturbative expression of the dressing
phase has been proposed for strings \cite{Beisert:2006ib}. It
agrees with the known string theory data and respects
crossing-symmetry. The phase factor has also been proposed for the
weakly-coupled $\mathcal{N}=4$ gauge theory and further evidence
was found that it is indeed related to the dressing factor from
string theory by analytic continuation
\cite{Eden:2006rx}\nocite{Beisert:2006ez}-\cite{Beisert:2007hz}.

\smallskip

The two-particle S-matrix for superstrings on $AdS_{5}\times
S^{5}$ was recently determined using the symmetry invariance in
\cite{Arutyunov:2006yd}. This two-body S-matrix obeys the standard
properties:
\begin{description}
  \item[Yang-Baxter Equation] $S_{23}S_{13}S_{12}=S_{12}S_{13}S_{23}$
  \item[Unitarity Condition]
  $S_{12}(p_{1},p_{2})S_{21}(p_{2},p_{1})=\mathbb{I}$
  \item[Hermitian Analyticity]
  $S_{21}(p_{2},p_{1})^{\dag}=S_{12}(p_{1},p_{2})$
  \item[Crossing Symmetry]
  $\mathcal{C}^{-1}_{1}S_{12}^{t_{1}}(p_{1},p_{2})\mathcal{C}_{1}S_{12}(-p_{1},p_{2})=\mathbb{I}$,
\end{description}
where $\mathcal{C}$ is the charge conjugation matrix.

\smallskip

In general one  encounters states with more than two excitations
and hence one would also need a multi-particle S-matrix. However,
the two-particle S-matrix contains all the relevant information if
one assumes integrability. Scattering in integrable models
preserves the number of particles and the set of their on-shell
momenta \cite{Zamolodchikov:1978xm}. In other words, there is no
particle production and in the scattering process the particle
momenta are merely exchanged. But, more importantly, these models
admit factorization of the S-matrix, i.e. any multi-particle
S-matrix, describing some scattering process, factorizes in a
product of two-particle $S$-matrices. Note, nonetheless, that the
string S-matrix we are considering, does not depend on the
difference of rapidities, as is normally the case in relativistic
two-dimensional integrable models possessing Lorentz symmetry. The
factorized scattering is an extremely useful property, since it
allows one to obtain the spectrum of a model from the two-particle
S-matrix only.

\smallskip

Let us now explain how to derive the energy spectrum from the two
particle S-matrix in the string theoretic picture and how the
Bethe equations come into play. Consider creation operators
$A^{\dag}_{M}$ and annihilation operators $A_{M}$. The algebra
these operators satisfy is the so-called Faddeev-Zamolodchikov
(ZF) algebra \cite{Zamolodchikov:1978xm}, \cite{Faddeev:1980zy}:
\begin{eqnarray}
A_{1}A_{2}=S_{12}A_{2}A_{1},\quad
A^{\dag}_{1}A_{2}^{\dag}=A^{\dag}_{2}A_{1}^{\dag}S_{12},\quad
A_{1}A^{\dag}_{2}=A_{2}^{\dag}S_{12}A_{1} + \delta_{12},
\end{eqnarray}
where $S_{12}$ is the two-particle S-matrix and $\delta_{12}$ is
the delta function depending on the difference of the momenta of
the scattering particles. One can recognize that the standard
properties of the two-particle S-matrix described above follow by
requiring consistency of the ZF algebra relations. Asymptotic
states are then constructed by acting with creation operators on
the vacuum $|0\rangle$. A generic state, consisting of excitations
with momentum $p_{i}$, will be of the form:
\begin{eqnarray}
A_{M_{1}}^{\dag}(p_{1})\ldots A_{M_{N}}^{\dag}(p_{N})|0\rangle.
\end{eqnarray}
The Hamiltonian is given by one of the central charges of the
symmetry algebra and hence, the dispersion relation is known
\cite{Beisert:2004hm}. From this, one can find the energy of such
a state \cite{Beisert:2004hm}:
\begin{eqnarray}
E = \sum_{i=1}^{N}\sqrt{1+16g^{2}\sin^{2}(\frac{1}{2}p_{i})}.
\end{eqnarray}
This holds for any values of the momenta $p_{i}$. However, since
we are dealing with  closed strings, we have to impose periodicity
on the wave function of  world-sheet excitations. This requirement
puts a restriction on the momenta in the form of a set of
equations, usually referred to as (nested) Bethe equations. For
the model in question this has been done recently by applying the
algebraic Bethe ansatz approach \cite{Martins:2007hb}.

\smallskip

In this paper we will rederive the nested Bethe equations by using
the coordinate Bethe ansatz in a way similar to
\cite{Beisert:2005tm}. First, we discuss the string S-matrix,
present the equations obtained in \cite{Martins:2007hb} and
briefly comment on how they were derived. Then we explain how the
nested Bethe ansatz works, followed by a more detailed discussion
on the involved calculations. We will also point out where the
calculations differ from \cite{Beisert:2005tm}. These results will
be used to obtain the Bethe equations, which coincide with the
ones found in \cite{Martins:2007hb}\footnote{There is a subtle
sign issue which is discussed in the next section.}. We will also
compare the obtained equations to the ones proposed in
\cite{Beisert:2005fw}. Finally, as a byproduct of our procedure we
also obtain the explicit form of the Bethe wave function.

\section{The S-Matrix and Algebraic Bethe Ansatz}

By demanding compatibility of the S-matrix describing world-sheet
scattering, with centrally extended $\mathfrak{su}(2|2)$ symmetry,
one can determine the S-matrix for strings on $AdS_{5}\times
S^{5}$ up to a phase factor \cite{Arutyunov:2006yd}. We will
consider this S-matrix:
\begin{eqnarray}
\widehat{S}^{\mathrm{I}}_{12}=S_{12}^{\mathrm{string}}(p_{1},p_{2}).
\end{eqnarray}
It acts according to the ZF algebra:
\begin{eqnarray}\label{SinZF}
\widehat{S}^{\mathrm{I}}_{12}\cdot
A^{\dag}_{M_{1}}(p_{1})A^{\dag}_{M_{2}}(p_{2}) =
S_{M_{1}M_{2}}^{N_{1}N_{2}}
A^{\dag}_{N_{2}}(p_{2})A^{\dag}_{N_{1}}(p_{1}),
\end{eqnarray}
where the sum convention is used. Let us write the components of
the S-matrix in the following way:
\begin{eqnarray}
S_{12}^{\mathrm{I}}|A^{\dag}_{a}(p_{1})A^{\dag}_{b}(p_{2})\rangle
&=& A|A^{\dag}_{\{a}(p_{2})A^{\dag}_{b\}}(p_{1})\rangle +
B|A^{\dag}_{[a}(p_{2})A^{\dag}_{b]}(p_{1})\rangle\nonumber\\
&& +
\frac{1}{2}C\epsilon_{ab}\epsilon^{\alpha\beta}|A^{\dag}_{\alpha}(p_{2})A^{\dag}_{\beta}(p_{1})\rangle\nonumber\\
S_{12}^{\mathrm{I}}|A^{\dag}_{\alpha}(p_{1})A^{\dag}_{\beta}(p_{2})\rangle
&=& D|A^{\dag}_{\{\alpha}(p_{2})A^{\dag}_{\beta\}}(p_{1})\rangle +
E|A^{\dag}_{[\alpha}(p_{2})A^{\dag}_{\beta]}(p_{1})\rangle\nonumber\\
&&+
\frac{1}{2}F\epsilon_{\alpha\beta}\epsilon^{ab}|A^{\dag}_{a}(p_{2})A^{\dag}_{b}(p_{1})\rangle\nonumber\\
S_{12}^{\mathrm{I}}|A^{\dag}_{a}(p_{1})A^{\dag}_{\beta}(p_{2})\rangle
&=& G|A^{\dag}_{\beta}(p_{2})A^{\dag}_{a}(p_{1})\rangle +
H|A^{\dag}_{a}(p_{2})A^{\dag}_{\beta}(p_{1})\rangle \nonumber\\
S_{12}^{\mathrm{I}}|A^{\dag}_{\alpha}(p_{1})A^{\dag}_{b}(p_{2})\rangle
&=& K|A^{\dag}_{\alpha}(p_{2})A^{\dag}_{b}(p_{1})\rangle +
L|A^{\dag}_{b}(p_{2})A^{\dag}_{\alpha}(p_{1})\rangle.
\end{eqnarray}
We will use the convention that the index $M=1,2,3,4$ runs through
both bosonic and fermionic indices. The bosonic indices will be
labelled $a,b=1,2$ and the fermionic indices will be labelled
$\alpha,\beta=3,4$. The coefficients describing this scattering
are easily seen from (\ref{SinZF}) to be:
\begin{eqnarray}\label{AFZSmatrix}
\begin{array}{lll}
  A=a_{1}(p_{1},p_{2}) & & F=2a_{7}(p_{1},p_{2}) \nonumber\\
  B=-(a_{1}+2a_{2})(p_{1},p_{2}) & &G=a_{5}(p_{1},p_{2})\nonumber \\
  C=2a_{8}(p_{1},p_{2}) & & H=a_{10}(p_{1},p_{2}) \nonumber\\
  D=a_{3}(p_{1},p_{2}) & & K=a_{9}(p_{1},p_{2}) \nonumber\\
  E=-(a_{3}+2a_{4})(p_{1},p_{2}) & & L=a_{6}(p_{1},p_{2})
\end{array}
\end{eqnarray}
The explicit form of the factors $a_{i}$ is derived in
\cite{Arutyunov:2006yd} and is for convenience stated it in the
appendix.

It is instructive to compare this S-matrix to the one used in
\cite{Beisert:2005tm}.  The S-matrix derived in
\cite{Arutyunov:2006yd}, also describes the spin chain S-matrix,
by making a particular choice for the coefficients $a_{i}$, which
is given in the appendix. The relation of this spin chain
S-matrix, with the S-matrix derived by Beisert, $S^{B}$, in
\cite{Beisert:2005tm}, is given by complex conjugation
\begin{eqnarray}
S^{B}(p_{1},p_{2})=\overline{S}^{chain}(p_{1},p_{2}),
\end{eqnarray}
where $S^{chain}$ is the aforementioned chain version of the
S-matrix and we have chosen $\overline{x^{\pm}}=x^{\mp}$. This
relation is more convenient in our case than the one given in
\cite{Arutyunov:2006yd}:
\begin{eqnarray}
S^{B}(p_{1},p_{2})=P\mathcal{P}S^{chain}(p_{2},p_{1})\mathcal{P},
\end{eqnarray}
where $P$ and $\mathcal{P}$ are permutation and graded permutation
respectively. In the latter case, the identification of the
coefficients with $A,B$ etc. is a little less straightforward.

By imposing periodicity on the discussed system, one derives
restrictions on the momenta $p_{i}$. The algebraic version of the
nested Bethe ansatz was recently applied \cite{Martins:2007hb} to
derive the equations describing this. This is done by transforming
the string S-matrix to Shastry's graded R-matrix, which makes that
one can apply results earlier derived for the Hubbard model
\cite{martins-1997},\cite{Ramos:1996us}. From this, the Bethe
equations for the string excitations are obtained and are given
by:
\begin{eqnarray}
e^{ip_{k}\left(-L+N-\frac{m_{1}^{(1)}}{2}-\frac{m_{1}^{(2)}}{2}\right)}
&=& e^{iP}\prod_{i=1,i \neq
k}^{N}S_{0}(p_{k},p_{i})\left[\frac{x_{i}^{-}-x_{k}^{+}}{x_{i}^{+}-x_{k}^{-}}\right]^{2}\times\nonumber\\
&&\times
\prod_{\alpha=1}^{2}\prod_{j=1}^{m_{1}^{(\alpha)}}\left[\frac{y_{j}^{(\alpha)}-x^{-}_{k}}{y_{j}^{(\alpha)}-x^{+}_{k}}\right]\nonumber\\
e^{i\frac{P}{2}}\prod_{i=1}^{N}\left[\frac{y_{j}^{(\alpha)}-x^{-}_{i}}{y_{j}^{(\alpha)}-x^{+}_{i}}\right]
&=&
\prod_{l=1}^{m_{2}^{(\alpha)}}\left[\frac{v^{(\alpha)}_{j}-w_{l}^{(\alpha)}+\frac{i}{2g}}{v^{(\alpha)}_{j}-w_{l}^{(\alpha)}-\frac{i}{2g}}\right]\nonumber\\
\prod_{j=1}^{m_{1}^{(\alpha)}}\left[\frac{w_{l}^{(\alpha)}-v^{(\alpha)}_{j}+\frac{i}{2g}}{w_{l}^{(\alpha)}-v^{(\alpha)}_{j}-\frac{i}{2g}}\right]
&=&\prod_{k=1,k\neq
l}^{m_{2}^{(\alpha)}}\left[\frac{w_{l}^{(\alpha)}-w_{k}^{(\alpha)}+\frac{i}{g}}{w_{l}^{(\alpha)}-w_{k}^{(\alpha)}-\frac{i}{g}}\right],
\end{eqnarray}
where $y$ and $w$ are auxiliary parameters and $y$ and $v$ are
related via:
\begin{eqnarray}
v = y + \frac{1}{y}.
\end{eqnarray}
As one would expect, these equations are very similar to the ones
describing the $\mathfrak{su}(2|2)$ dynamic spin chain
\cite{Beisert:2005tm}, however, when comparing to
\cite{Beisert:2005fw}, we see that there is a slight mismatch
$L\leftrightarrow -L $, which is probably due the ambiguity in the
Bethe ansatz as noted at the end of Section 3 from
\cite{Martins:2007hb}. In \cite{Beisert:2005tm}, the coordinate
Bethe ansatz is used to derive the Bethe equations. In the
following sections we will make the comparison explicit, i.e. we
will rederive these equations by using the coordinate Bethe
ansatz. However, we will set up our ansatz in such a way that the
sign in front of $L$ will coincide with the one in
\cite{Beisert:2005fw}.

Let us conclude this section by briefly discussing the difference
between the algebraic and the coordinate Bethe ansatz approaches.
In the algebraic case, one considers the monodromy matrix of the
system. The starting point is to choose a particular state which
is annihilated by the lower triangular part of the monodromy
matrix. Then, from the upper triangular part, one can find
creation operators. These operators are used to construct
eigenstates of the trace of the monodromy (transfer matrix). From
this construction one obtains the Bethe equations.

In the coordinate Bethe ansatz, one makes an ansatz for the wave
function directly from the creation operators of the ZF algebra
acting on a vacuum. Then one imposes periodicity on this wave
function, which leads to the Bethe equations.

\section{Procedure}

In this section we will briefly discuss how the method of the
coordinate Bethe ansatz will be applied here. Most of the
calculational details will be treated in the next sections. The
nested Bethe ansatz was first introduced in a seminal paper
written by Yang \cite{Yang:1967bm}. We will mostly follow
\cite{Beisert:2005fw} and \cite{Beisert:2005tm}.

The problem one wishes to solve is how to impose the periodicity
condition on the wave function of the world-sheet excitations.
This is needed since we are dealing with (non-interacting) closed
strings of length described by a parameter $L$. Thus, the wave
functions corresponding to world-sheet excitations should be
$L$-periodic. The equations that capture this are called the Bethe
equations.

Let us introduce some notation. The different asymptotic string
states are built out of the ZF oscillators $A_{i}^{\dag}(p_{k})$
acting on a vacuum $|0\rangle$. Let us now consider the coordinate
space with coordinates $\sigma$ and suppose we create a state by
using $K^{\mathrm{I}}$ creation operators.

Consider the case $\sigma_{1}\gg\ldots\gg\sigma_{K^{\mathrm{I}}}$.
In this case, the excitations are far apart, which means that we
neglect the interaction between them. Consider a creation operator
$A^{\dag}_{M}(\sigma)$, which creates a particle with index $M$ at
position $\sigma$. By definition, the state
$|A^{\dag}_{M_{1}}(p_{1})\ldots
A^{\dag}_{M_{K^{\mathrm{I}}}}(p_{K^{\mathrm{I}}})\rangle:=A^{\dag}_{M_{1}}(p_{1})\ldots
A^{\dag}_{M_{K^{\mathrm{I}}}}(p_{K^{\mathrm{I}}})|0\rangle$
describes $K^{\mathrm{I}}$ particles such that the particle with
momentum $p_{i}$ is to the left of the particle with momentum
$p_{i+1}$. In other words, we have the identification:
\begin{eqnarray}\label{ZFgen}
&&|A^{\dag}_{M_{1}}(p_{1})\ldots A^{\dag}_{M_{K^{\mathrm{I}}}}(\sigma_{K^{\mathrm{I}}})\rangle=\\ &&\quad \int_{\sigma_{1}\gg\ldots\gg\sigma_{K^{\mathrm{I}}}}d\sigma_{1}\ldots d\sigma_{K^{\mathrm{I}}}e^{-i\sum_{j=1}^{K^{\mathrm{I}}}p_{j}\sigma_{j}}A^{\dag}_{M_{1}}(\sigma_{1})\ldots A^{\dag}_{M_{K^{\mathrm{I}}}}(\sigma_{K^{\mathrm{I}}})|0\rangle\nonumber.
\end{eqnarray}
The ansatz for the wave function in this sector is:
\begin{eqnarray}
\Phi(p_{1},\ldots,p_{K^{\mathrm{I}}})= \chi_{M_{1}\ldots
M_{K^{\mathrm{I}}}
}(p_{1},\ldots,p_{K^{\mathrm{I}}})|A^{\dag}_{M_{1}}(p_{1})\ldots
A^{\dag}_{M_{K^{\mathrm{I}}}}(\sigma_{K^{\mathrm{I}}})\rangle,
\end{eqnarray}
where the indices are summed over. More generally, if $Q$ is a
permutation of the numbers $(1,\ldots,K^{\mathrm{I}})$, then in
the sector where
$\sigma_{Q_{1}}\gg\ldots\gg\sigma_{Q_{K^{\mathrm{I}}}}$, we make a
similar ansatz:
\begin{eqnarray}\label{BaLvl1Coef}
\Phi_{Q}(q_{1},\ldots,q_{K^{\mathrm{I}}})=\chi_{Q;N_{1}\ldots
N_{K^{\mathrm{I}}}}(q_{1},\ldots,q_{K^{\mathrm{I}}})
A^{\dag}_{N_{1}}(q_{1})\ldots A^{\dag}_{N_{K^{\mathrm{I}
}}}(q_{K^{\mathrm{I}}})|0\rangle.
\end{eqnarray}
Note that we can also just see this as a wave function in the
sector $\sigma_{1}\gg\ldots\gg\sigma_{K^{\mathrm{I}}}$ with
permuted momenta, by just a simple relabelling of the integration
variables in (\ref{ZFgen}). The region where there is interaction
between the excitations links the different sectors and the
relation between them is, by definition, given by the S-matrix.
Thus, if we for example consider $Q=(12)$, then we obtain the
following relation:
\begin{eqnarray}
\chi_{12;N_{1}N_{2}\ldots}(p_{2},p_{1},\ldots,p_{K^{\mathrm{I}}})=
S^{M_{2}M_{1}}_{N_{1}N_{2}}\chi_{M_{1}M_{2}\ldots}(p_{1},p_{2},\ldots,p_{K^{\mathrm{I}}}).
\end{eqnarray}
More specifically, by the above relation, we can extend the
asymptotic state in the region
$\sigma_{1}\gg\ldots\gg\sigma_{K^{\mathrm{I}}}$ to the entire
string in a unique way. The complete wave function for this
asymptotic state will be given by:
\begin{eqnarray}
\chi_{M_{1}\ldots
M_{K^{\mathrm{I}}}}(p_{1},\ldots,p_{K^{\mathrm{I}}})\sum_{P\in
S_{K^{\mathrm{I}}}}\hat{S}_{P}|A_{M_{1}}^{\dag}(p_{1})\ldots
A_{M_{\mathrm{I}}}^{\dag}(p_{K^{\mathrm{I}}})\rangle^{\mathrm{I}}+\
\textrm{non-asymp},
\end{eqnarray}
where the sum runs over all permutations of $\{1,\ldots,K^{\mathrm{I}}\}$.
The periodicity condition is now formulated on the Fourier components of (\ref{ZFgen}) by demanding that the wave function is invariant under:
\begin{eqnarray}\label{period}
(\sigma_{1},\ldots,\sigma_{K^{\mathrm{I}}})&\rightarrow&(\sigma_{K^{\mathrm{I}}}+L,\sigma_{1},\sigma_{2},\ldots,\sigma_{K^{\mathrm{I}}-1})\nonumber\\
&\rightarrow&(\sigma_{K^{\mathrm{I}}-k}+L,\ldots,\sigma_{K^{\mathrm{I}}}+L,\sigma_{1},\sigma_{2},\ldots,\sigma_{K^{\mathrm{I}}-k-1})
\end{eqnarray}
for all $k\in\{1,\ldots,K^{\mathrm{I}}\}$. When we make ansatz
(\ref{BaLvl1Coef}), the periodicity condition (\ref{period}), for
any $k$, is given by:
\begin{eqnarray}
e^{-ip_{k}L}\hat{S}^{\mathrm{I}}_{(k,k+1)}\ldots\hat{S}^{\mathrm{I}}_{(k,K^{\mathrm{I}})}\hat{S}^{\mathrm{I}}_{(k,1)}\ldots\hat{S}^{\mathrm{I}}_{(k,k-1)}\Phi(p)=\Phi(p),
\end{eqnarray}
or written out explicitly:
\begin{eqnarray}\label{BAE}
&&e^{-ip_{K^{\mathrm{I}}}L}S^{\lambda_{1}\mu_{1}}_{M_{1}M_{2}}(p_{k},p_{k+1})S^{\lambda_{2}\mu_{2}}_{\mu_{1}M_{3}}(p_{k},p_{k+2})\ldots
S^{\lambda_{K^{\mathrm{I}}}\lambda_{K^{\mathrm{I}}-1}}_{\mu_{K^{\mathrm{I}}}M_{K^{\mathrm{I}}}}(p_{k},p_{k-1})\chi_{\lambda_{1}\ldots \lambda_{K^{\mathrm{I}}}}\nonumber\\
&&\qquad= \chi_{M_{1}\ldots M_{K^{\mathrm{I}}}}.
\end{eqnarray}
The term $S(p_{k},p_{k})$ is of course absent in the above
product.

We are now left with solving this equation for the coefficients
$\chi$. This can be solved by making use of auxiliary systems that
allow for additional Bethe ans\"atze. This is the nesting.
Equation (\ref{BAE}) can be seen as a matrix equation:
\begin{eqnarray}
T_{k,M_{1}\ldots M_{K^{\mathrm{I}}}}^{\lambda_{1}\ldots
\lambda_{K^{\mathrm{I}}}}\chi_{\lambda_{1}\ldots
\lambda_{K^{\mathrm{I}}}}= \chi_{M_{1}\ldots M_{K^{\mathrm{I}}}}.
\end{eqnarray}
The nesting procedure means that we will find the eigenvectors of
the matrix operator $T$ in steps. Finally, note that we have
$K^{\mathrm{I}}$ of these equations, but from the Yang-Baxter
equation it is easily verified that the different matrices $T$ all
commute and hence can be diagonalized simultaneously.

\begin{figure}
\begin{center}
\psfrag{Tag1}[][l]{} \psfrag{Tag2}[][l]{$A^{\dag}_{1}$}
\psfrag{Tag3}[][l]{$A^{\dag}_{3}$}
\psfrag{Tag4}[][l]{$A^{\dag}_{4}$}
\psfrag{Tag5}[][l]{$K^{\mathrm{I}}$ excitations on a length $L$
string} \psfrag{Tag6}[][l]{$K^{\mathrm{II}}$ excitations}
\psfrag{Tag7}[][l]{$K^{\mathrm{III}}$ excitations}
\includegraphics{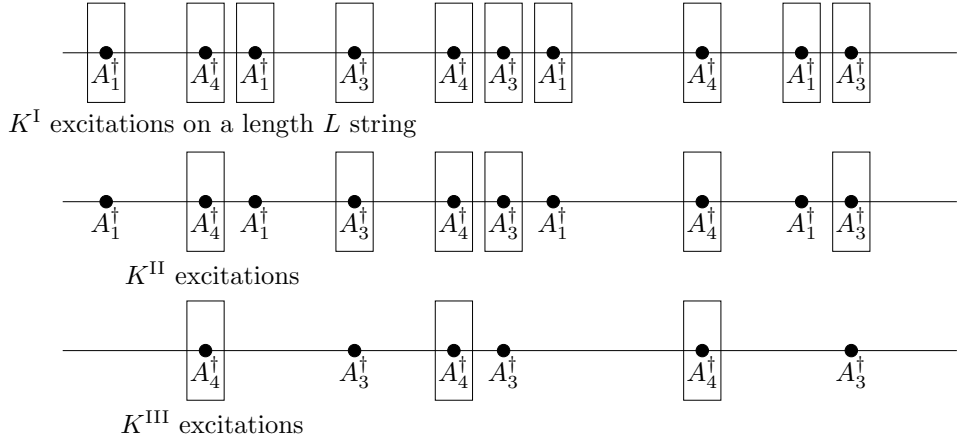}
\end{center}
\caption{Schematic representation of the different levels of the
nested Bethe ansatz. At each level, the plain dots represent the
operators forming the vacuum and the boxes stand for the operators
that are creation operators. }\label{FigNBA}
\end{figure}

The idea is that we work in different steps or levels to
diagonalize these matrices. This is done by considering auxiliary
periodic systems. At each level we specify a ``new" vacuum and
``new" creation operators. The procedure is illustrated in the
Figure \ref{FigNBA}. Each time a box indicates which operators are
considered as creation operators and the operators without the box
are considered background or vacuum. The end result is that we
find appropriate coefficient $\chi$ which solves equation
(\ref{BAE}) and hence we obtain the explicit wave functions of the
system.

We start with the first level. The wave function for this level is
given by the product of ZF generators:
\begin{eqnarray}
|A^{\dag}_{M_{1}}(p_{1})\ldots A^{\dag}_{M_{K^{\mathrm{I}}}}(p_{K^{\mathrm{I}}})\rangle^{\mathrm{I}}:=
|A^{\dag}_{M_{1}}(p_{1})\ldots A^{\dag}_{M_{K^{\mathrm{I}}}}(p_{K^{\mathrm{I}}})\rangle.
\end{eqnarray}
In this level, we have $K^{\mathrm{I}}$
excitations, or creation operators. Since we assume integrability,
we know that this number is conserved.

For the next level, we define the first auxiliary system. This
system is just a chain with $K^{\mathrm{I}}$ sites. One has to
define what one considers as the vacuum state and what operators
are to be considered as excitations. This is analogous to, for
example, the Heisenberg spin chain where one can take all spins
down to be the vacuum and one considers spins up as excitations.
The choice made for the reference state at the second level is:
\begin{eqnarray}
|0\rangle^{\mathrm{II}}=|A^{\dag}_{1}(p_{1})\ldots
A^{\dag}_{1}(p_{K^{\mathrm{I}}})\rangle^{\mathrm{I}},
\end{eqnarray}
and all the other creation operators are considered to be creation
operators on this new vacuum. This is shown in the second line of
Figure \ref{FigNBA}. In this section we exclude the $A_{2}^{\dag}$
excitations from the discussion since there is a subtle point
about them which will be treated in the next section. However, for
the understanding of the process, the absence of $A^{\dag}_{2}$
plays no role. Now, one makes a second Bethe ansatz for this
level. In this ansatz we will encounter additional parameters $y$,
which will play the role of the momenta at this level. For one
excitation, consisting of an $A^{\dag}_{\alpha}$, the ansatz takes
the form:
\begin{eqnarray}
|A^{\dag}_{\alpha}\rangle^{\mathrm{II}}=\sum_{k=1}^{K^{\mathrm{I}}}\Psi_{k}^{(1)}(y)|A^{\dag}_{1}(p_{1})\ldots
A^{\dag}_{\alpha}(p_{k})\ldots
A^{\dag}_{1}(p_{K^{\mathrm{I}}})\rangle^{\mathrm{I}}.
\end{eqnarray}
This is just like a sum of plane waves. The way to determine the
coefficients $\Psi_{k}(y)$ is to impose compatibility with the
S-matrix, i.e.:
\begin{eqnarray}
S^{\mathrm{I}}_{(k,l)}|A^{\dag}_{\alpha}\rangle^{\mathrm{II}}
=S^{\mathrm{I,I}}_{k,l}(p_{k},p_{l})|A^{\dag}_{\alpha}\rangle^{\mathrm{II}}_{(k,l)},
\end{eqnarray}
where $|A^{\dag}_{\alpha}\rangle^{\mathrm{II}}_{(k,l)}$ is
$|A^{\dag}_{\alpha}\rangle^{\mathrm{II}}$ with the momenta $p_{k}$
and $p_{l}$ interchanged and
\newline\noindent
$S^{\mathrm{I,I}}_{(k,l)}(p_{k},p_{l})$ is a phase factor. This is
a natural condition to impose, since this basically implies that
the state, obtained in this way, is an eigenstate of the matrix
$T$.

What the explicit form of $\Psi_{k}(y)$ is and how to deal with
more than one excitation will be treated in the next section.
However, the bottom line of this procedure is that we are one step
closer to imposing periodicity and we now need to consider one
creation operator less at the cost of introducing extra momenta
$y$. We call the number of excitations at this level
$K^{\mathrm{II}}$, this can be interpreted as number of fermions
in this system. In order to impose periodicity at this level, we
again need to introduce an additional auxiliary system.

We proceed in a similar way and choose the reference state in the
next level as
\begin{eqnarray}\label{VacIII}
|0\rangle^{\mathrm{III}}=|A^{\dag}_{3}\ldots
A^{\dag}_{3}\rangle^{\mathrm{II}}
\end{eqnarray}
and one only considers $A^{\dag}_{4}$ as a creation operator. The
Bethe ansatz made this time for a single excitation is of the same
form as in the previous level:
\begin{eqnarray}
|A^{\dag}_{4}\rangle^{\mathrm{III}}(w)=\sum_{k=1}^{K^{\mathrm{II}}}\Psi_{k}^{(2)}(w)|A^{\dag}_{3}(y_{1})\ldots
A^{\dag}_{4}(y_{k})\ldots
A^{\dag}_{3}(y_{K^{\mathrm{II}}})\rangle^{\mathrm{II}}.
\end{eqnarray}
One can now determine the coefficients by imposing compatibility
with the level II S-matrix, which roughly describes the scattering
of level II wave functions. The system is now reduced to just one
type of creation operators of which there are $K^{\mathrm{III}}$.
This means that the wave function is fixed by giving the three
different numbers of creation operators
$K^{\mathrm{I}},K^{\mathrm{II}},K^{\mathrm{III}}$ and three sets
of momenta, $\{p,y,w\}$.

By imposing the periodicity condition, one can derive the Bethe
equations for the system. Note that periodicity is present in all
three levels, which will give three sets of equations. The first
one will, of course, correspond to the eigenvalues of the matrix
equation (\ref{BAE}). The other ones will put restrictions on the
auxiliary momenta $y,w$. They will be derived in Section 5.

\begin{figure}
\begin{center}
\psfrag{Tag1}[][l]{$S$}
\includegraphics{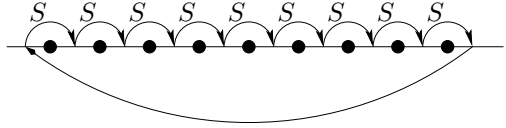}
\end{center}
\caption{Schematic representation of the periodicity condition.
}\label{FigPeriodBA}
\end{figure}

The Bethe equations can be seen as scattering, via the relevant
S-matrices, a creation operator around the string at the different
levels. Each time one scatters two operators, the wave function
picks up a phase factor. When the operator is back at its original
position, the wave function should be unchanged (up to a phase
factor). This is schematically depicted in Figure
\ref{FigPeriodBA}.
This amounts to the following Bethe equations:
\begin{eqnarray}\label{RealBAE}
e^{iL_{A,k}}=
{\prod_{B=\mathrm{I}}^{\mathrm{III}}\prod_{l=1}^{K^{B}}}
S^{BA}(x^{B}_{l},x^{A}_{k}),\quad (B,l)\neq (A,k)
\end{eqnarray}
where $A,B$ denote the different levels and, roughly, $S^{AB}$ is
the S-matrix describing how an excitation at level $B$ is
scattered with an excitation of level $A$. Moreover,
$e^{iL_{\mathrm{I},k}}=e^{iLp_{k}}$ and
$e^{iL_{\mathrm{II},k}}=e^{iL_{\mathrm{III},k}}=1$ are phases
depending on the level that is considered. This formula will be
derived later on.

The phase $e^{iLp_{k}}$ is dependent on the length of the string,
$L$. When working in the uniform light-cone gauge
\cite{Arutyunov:2005hd}, \cite{Arutyunov:2004yx} one can express
the length in terms of the conserved $U(1)$ charge $J$ of the
string: $L=J$.

\section{Levels of the S-Matrix}

In this section we will derive the explicit form of the wave
functions as well as the factors of the S-matrix corresponding to
the different levels.

\subsection{$S^{\mathrm{I,I}}$}

Recall that the level $\mathrm{II}$ reference state is given by
$|0\rangle^{\mathrm{II}} = |A^{\dag}_{1}(p_{1})\ldots
A^{\dag}_{1}(p_{K^{\mathrm{I}}})\rangle^{\mathrm{I}}$. Since we
assumed integrability, we only need to consider the action of a
two particle S-matrix. As is easily seen from (\ref{AFZSmatrix}),
the S-matrix acts trivially on the reference state at this level:
\begin{eqnarray}
\hat{S}_{(k,l)}^{\mathrm{I}}|0\rangle^{\mathrm{II}}&=:&S^{\mathrm{I,I}}(p_{k},p_{l})|0\rangle^{\mathrm{II}}_{(k,l)}\nonumber\\
&=&A(p_{k},p_{l})|0\rangle^{\mathrm{II}}_{(k,l)}\nonumber\\
&=&S_{0}(p_{k},p_{l})\left[\frac{x_{l}^{-}-x_{k}^{+}}{x_{l}^{+}-x_{k}^{-}}\right]\frac{e^{i\frac{p_{l}}{2}}}{e^{i\frac{p_{k}}{2}}}|0\rangle^{\mathrm{II}}_{(k,l)}
\end{eqnarray}
where $(k,l)$ are the two particles that scatter and
$|0\rangle^{\mathrm{II}}_{(k,l)}$ is $|0\rangle^{\mathrm{II}}$
with $p_{k}$ and $p_{l}$ interchanged. The factor $S_{0}$ is the
undetermined scalar phase of the S-matrix is given in the
appendix.

\subsection{Propagation and $S^{\mathrm{II,I}}$}

The next step is to consider excitations in this level. Let us
start by considering a single excitation and see how this
``propagates" in the vacuum $|0\rangle^{\mathrm{II}}$. From
(\ref{AFZSmatrix}), it is easily seen that an insertion of
$A^{\dag}_{2}$, in a sea of $ A^{\dag}_{1}$ fields can ``decay"
into the operators $A^{\dag}_{3}$ and $A^{\dag}_{4}$. Hence, the
$A^{\dag}_{2}$ behaves like a double excitation with respect to
this reference state and we do not consider it here. A generic
one-excitation state is now given by:
\begin{eqnarray}
|A^{\dag}_{\alpha}\rangle^{\mathrm{II}} =
\sum_{k=1}^{K^{\mathrm{I}}}\Psi_{k}^{(1)}|
A^{\dag}_{1}(p_{1})\ldots A^{\dag}_{\alpha}(p_{k})\ldots
A^{\dag}_{1}(p_{K^{\mathrm{I}}})\rangle^{\mathrm{I}}.
\end{eqnarray}
The ansatz made for the coefficient is the following:
\begin{eqnarray}
\Psi_{k}^{(1)}=f(x_{k})\prod_{l=1}^{k-1}S^{\mathrm{II},\mathrm{I}}(x_{l}).
\end{eqnarray}
The terms in the above expression can be viewed as a factor
obtained by permuting the excitation with the background field
($S^{\mathrm{II},\mathrm{I}}$), together with a factor for the
combination of the excitation with the background field at
position $k$ $(f(x_{k}))$. As discussed in the previous section,
one imposes compatibility with the level I S-matrix:
\begin{eqnarray}
S^{\mathrm{I}}_{(k,l)}|A^{\dag}_{\alpha}\rangle^{\mathrm{II}}
=S^{\mathrm{I,I}}_{k,l}(p_{k},p_{l})|A^{\dag}_{\alpha}\rangle^{\mathrm{II}}_{(k,l)},
\end{eqnarray}
where $|A^{\dag}_{\alpha}\rangle^{\mathrm{II}}_{(k,l)}$ is
$|A^{\dag}_{\alpha}\rangle^{\mathrm{II}}$ with the momenta $p_{k}$
and $p_{l}$ interchanged.

To explicitly solve the functions $f$ and
$S^{\mathrm{II},\mathrm{I}}$, it is enough to consider a chain
with only two sites:
\begin{eqnarray}
|A^{\dag}_{\alpha}\rangle^{\mathrm{II}}
&=&f(x_{1})|A^{\dag}_{\alpha}(p_{1})
A^{\dag}_{1}(p_{2})\rangle^{\mathrm{I}} +
f(x_{2})S^{\mathrm{II,I}}(x_{1})| A^{\dag}_{1}(p_{1})A^{\dag}_{\alpha}(p_{2})\rangle^{\mathrm{I}}\nonumber\\
|A^{\dag}_{\alpha}\rangle^{\mathrm{II}}_{(1,2)}
&=&f(x_{2})|A^{\dag}_{\alpha}(p_{2})
A^{\dag}_{1}(p_{1})\rangle^{\mathrm{I}} +
f(x_{1})S^{\mathrm{II,I}}(x_{2})|
A^{\dag}_{1}(p_{2})A^{\dag}_{\alpha}(p_{1})\rangle^{\mathrm{I}}.
\end{eqnarray}
Written out, the above compatibility condition gives the following
equations:
\begin{eqnarray}
f(x_{1})K + f(x_{2})S^{\mathrm{II,I}}(x_{1})G &=&
f(x_{2})A(x_{1},x_{2})\nonumber\\
f(x_{1})L + f(x_{2})S^{\mathrm{II,I}}(x_{1})H &=&
f(x_{1})S^{\mathrm{II,I}}(x_{2})A(x_{1},x_{2}).
\end{eqnarray}
Now, one uses the first equation to solve for
$S^{\mathrm{II,I}}(x)$ in terms of $x^{\pm}$ and $f(x)$. This
result can then be used, together with the second equation, to
solve for $f(x_{1})$ in terms of $x_{1}^{-},x_{2}^{-}$ and
$f(x_{2})$. By differentiating this expression with respect to
$x_{2}^{-}$, one can solve for $f(x_{2})$ only in terms of
$x_{2}^{-}$. Doing this, one obtains
\begin{eqnarray}
f(x_{k})&=&e^{-i\frac{p_{k}}{2}}\frac{\eta(p_{k})y}{y-x_{k}^{-}},
\end{eqnarray}
where $y$ is an integration constant. The constant $y$ will play
the role of a pseudo-momentum and we will explicitly include it in
our notation from now on. The explicit solution for
$S^{\mathrm{II,I}}$ is easily obtained by using the above form of
$f$. The complete solution is given by:
\begin{eqnarray}
S^{\mathrm{II,I}}(y,x_{k})&=&e^{-i\frac{p_{k}}{2}}\frac{y-x_{k}^{+}}{y-x_{k}^{-}}\nonumber\\
f(y,x_{k})&=&\eta(p_{k})e^{-i\frac{p_{k}}{2}}\frac{y}{y-x_{k}^{-}}.
\end{eqnarray}

\subsection{Scattering and $S^{\mathrm{II,II}}$}

By using similar techniques, one also solves for the two
excitation case. However, here one encounters additional degrees
of freedom, which are dealt with by introducing an additional
level of the S-matrix.

The natural ansatz to make for a two-excitation state is the
superposition of two one-magnon states:
\begin{eqnarray}
&&|A^{\dag}_{\alpha}(y_{1})A^{\dag}_{\beta}(y_{2})\rangle^{\mathrm{II}}
=\\
&&\qquad\sum_{k<l=1}^{K^{\mathrm{I}}}\Psi_{k}^{(1)}(y_{1})\Psi_{l}^{(1)}(y_{2})|
A^{\dag}_{1}(p_{1})\ldots A^{\dag}_{\alpha}(p_{k})\ldots
A^{\dag}_{\beta}(p_{l})\ldots
A^{\dag}_{1}(p_{K^{\mathrm{I}}})\rangle^{\mathrm{I}}\nonumber.
\end{eqnarray}
It is easy to see that this solves the compatibility condition
mentioned above if the two excitations are not neighbors. The
additional freedom can be seen from the above formula. In this
ansatz, we always have $y_{1}$ to the left of $y_{2}$, so this is
very similar to the normal Bethe ansatz for a spin chain. In
analogy to this, we introduce a second level S-matrix,
$S^{\mathrm{II}}$, which deals with interchanging $y_{1}$ and
$y_{2}$. So a general two-excitation state
$|A^{\dag}_{\alpha}A^{\dag}_{\beta}\rangle^{\mathrm{II}}$,
consisting of $A^{\dag}_{\alpha}$ and $A^{\dag}_{\beta}$, will be
of the form:
\begin{eqnarray}
|A^{\dag}_{\alpha}A^{\dag}_{\beta}\rangle^{\mathrm{II}}=|A^{\dag}_{\alpha}(y_{1})A^{\dag}_{\beta}(y_{2})\rangle^{\mathrm{II}}
+
S^{\mathrm{II}}_{12}(y_{1},y_{2})|A^{\dag}_{\alpha}(y_{1})A^{\dag}_{\beta}(y_{2})\rangle^{\mathrm{II}},
\end{eqnarray}
with
\begin{eqnarray}\label{ScatteringII}
S^{\mathrm{II}}_{12}(y_{1},y_{2})|A^{\dag}_{\alpha}(y_{1})A^{\dag}_{\beta}(y_{2})\rangle^{\mathrm{II}}&=&
M_{12}(y_{1},y_{2})|A^{\dag}_{\alpha}(y_{2})A^{\dag}_{\beta}(y_{1})\rangle^{\mathrm{II}}\nonumber\\
&&+
N_{12}(y_{1},y_{2})|A^{\dag}_{\beta}(y_{2})A^{\dag}_{\alpha}(y_{1})\rangle^{\mathrm{II}}.
\end{eqnarray}
Indeed, $S^{\mathrm{II}}$ acts like a S-matrix with respect to the
momenta $y_{i}$.

The $A^{\dag}_{2}$ field has to be included as well. Since this
behaves like a double excitation in the $ A^{\dag}_{1}$
background, we do not have the additional freedom corresponding to
interchanging the $y$s. Instead, one includes an additional factor
$f(y_{1},y_{2},x_{k})$, which occurs when two excitations reside
on the same position. This leads to the following ansatz:
\begin{eqnarray}
|A^{\dag}_{2}\rangle^{\mathrm{II}}
=\sum_{k=1}^{K^{\mathrm{I}}}\Psi_{k}^{(1)}(y_{1})\Psi_{k}^{(1)}(y_{2})f(y_{1},y_{2},x_{k})|
A^{\dag}_{1}(p_{1})\ldots A^{\dag}_{2}(p_{k})\ldots
A^{\dag}_{1}(p_{K^{\mathrm{I}}})\rangle^{\mathrm{I}}.
\end{eqnarray}
To sum it all up, a general two excitation state is given by:
\begin{eqnarray}
|2\rangle^{\mathrm{II}}=|A^{\dag}_{\alpha}(y_{1})A^{\dag}_{\beta}(y_{2})\rangle
+ \epsilon^{\alpha\beta}|A^{\dag}_{2}\rangle +
S_{12}^{\mathrm{II}}|A^{\dag}_{\alpha}(y_{1})A^{\dag}_{\beta}(y_{2})\rangle.
\end{eqnarray}
The explicit form of the second level S-matrix and the
two-excitation factor can be obtained by imposing compatibility
with the level $\mathrm{I}$ S-matrix. When we explicitly write
down the two-excitation state, we get:
\begin{eqnarray}\label{TwoExcState}
&&\ |2\rangle^{\mathrm{II}}=\\
&&\ \Phi(x_{1},x_{2})f(y_{1},x_{1})f(y_{2},x_{2})S^{\mathrm{II,I}}(y_{2},x_{1})|A^{\dag}_{\alpha}(p_{1})A^{\dag}_{\beta}(p_{2})\rangle^{\mathrm{I}}\nonumber\\
&&\ +f(y_{1},x_{1})f(y_{2},x_{1})f(y_{1},y_{2},x_{1})\epsilon^{\alpha\beta}|A^{\dag}_{2}(p_{1}) A^{\dag}_{1}(p_{2})\rangle^{\mathrm{I}}\nonumber\\
&&\ +f(y_{1},x_{2})f(y_{2},x_{2})S^{\mathrm{II,I}}(y_{1},x_{1})S^{\mathrm{II,I}}(y_{2},x_{1})f(y_{1},y_{2},x_{2})\epsilon^{\alpha\beta}|A^{\dag}_{1}(p_{1}) A^{\dag}_{2}(p_{2})\rangle^{\mathrm{I}}\nonumber\\
&&\ +\Phi(x_{1},x_{2})M(y_{1},y_{2})f(y_{2},x_{1})f(y_{1},x_{2})S^{\mathrm{II,I}}(y_{1},x_{1})|A^{\dag}_{\alpha}(p_{1})A^{\dag}_{\beta}(p_{2})\rangle^{\mathrm{I}}\nonumber\\
&&\
+\Phi(x_{1},x_{2})N(y_{1},y_{2})f(y_{2},x_{1})f(y_{1},x_{2})S^{\mathrm{II,I}}(y_{1},x_{1})|A^{\dag}_{\beta}(p_{1})A^{\dag}_{\alpha}(p_{2})\rangle^{\mathrm{I}}.\nonumber
\end{eqnarray}
When we write out the compatibility condition, we get four
equations corresponding to the different configurations in
(\ref{TwoExcState}). In order to make the equations not more
cumbersome than they already are, we introduce the short-hand
notation $f_{kl}:=f(y_{k},x_{l}),
S_{kl}:=S^{\mathrm{II,I}}(y_{k},x_{l}), M:=
M_{12}(y_{1},y_{2}),N:= N_{12}(y_{1},y_{2})$. The equations coming
from the configurations
$|A^{\dag}_{\alpha}A^{\dag}_{\beta}\rangle$ and
$|A^{\dag}_{\beta}A^{\dag}_{\alpha}\rangle$ are given by:
\begin{eqnarray}\label{TwoExcComp1}
\{f_{12}f_{21}S_{22} + M f_{22}f_{11}S_{12}\}\Phi_{21}A &=&
\{f_{11}f_{22}S_{21} + M
f_{21}f_{12}S_{11}\}\Phi_{12}\frac{D+E}{2}\nonumber\\
&&+ N f_{21}f_{12}S_{11}\Phi_{12}\frac{D-E}{2}\\
&&+ \left( -f_{11}f_{21}f_{121} + f_{12}f_{22}S_{11}S_{21}f_{122}
\right)\frac{C}{2}\nonumber
\end{eqnarray}
and
\begin{eqnarray}\label{TwoExcComp2}
Nf_{22}f_{11}S_{12}\Phi_{21}A &=& \{f_{11}f_{22}S_{21} + M
f_{21}f_{12}S_{11}\}\Phi_{12}\frac{D-E}{2}\nonumber\\
&&+ N f_{21}f_{12}S_{11}\Phi_{12}\frac{D+E}{2}\\
&&- \left( -f_{11}f_{21}f_{121} + f_{12}f_{22}S_{11}S_{21}f_{122}
\right)\frac{C}{2}.\nonumber
\end{eqnarray}
The equations coming from the double excitation $A^{\dag}_{2}$ are
easily see to be given by:
\begin{eqnarray}\label{TwoExcComp3}
f_{11}f_{21}S_{12}S_{22}f_{121}A &=& \{f_{11}f_{22}S_{21} + (M-N)
f_{21}f_{12}S_{11}\}\Phi_{12}\frac{F}{2}\\
&&+f_{11}f_{21}f_{121}\frac{A-B}{2} +
f_{12}f_{22}S_{11}S_{21}f_{122}\frac{A+B}{2}\nonumber
\end{eqnarray}
and
\begin{eqnarray}\label{TwoExcComp4}
f_{12}f_{22}f_{122}A &=& -\{f_{11}f_{22}S_{21} + (M-N)
f_{21}f_{12}S_{11}\}\Phi_{12}\frac{F}{2}\\
&&+f_{11}f_{21}f_{121}\frac{A+B}{2} +
f_{12}f_{22}S_{11}S_{21}f_{122}\frac{A-B}{2}.\nonumber
\end{eqnarray}
First, we add equations (\ref{TwoExcComp1}) and
(\ref{TwoExcComp2}), which yields:
\begin{eqnarray}
M+N=-1.
\end{eqnarray}
From adding (\ref{TwoExcComp3}) and (\ref{TwoExcComp4}), it can be
easily seen that $f(y_{1},y_{2},x_{k})$ must be of the form:
\begin{eqnarray}
f(y_{1},y_{2},x_{k})
=\frac{(y_{1}y_{2}-x_{k}^{+}x_{k}^{-})}{\eta(p_{k})^{2}}\frac{(x_{k}^{+}-x_{k}^{-})}{x_{k}^{-}}h(y_{1},y_{2}).
\end{eqnarray}
Finally, this leaves us to determine $M-N$ and the factor
$h(y_{1},y_{2})$. This is done by subtracting equations
(\ref{TwoExcComp1}) and (\ref{TwoExcComp2}) and by substraction of
equations (\ref{TwoExcComp3}) and (\ref{TwoExcComp4}). The
resulting equations can be solved analytically. First, one solves
for $M-N$, by eliminating $h(y_{1},y_{2})$. This yields a fraction
whose numerator and denominator are both polynomials in
$y_{1},y_{2}$, with coefficients depending on $x_{1},x_{2}$. By
doing a careful analysis of the numerator and denominator and by
comparing the two, using the relation (\ref{parameters}), one
obtains the following result:
\begin{eqnarray}
M-N&=&\frac{v_{1}-v_{2}+\frac{i}{g}}{v_{1}-v_{2}-\frac{i}{g}},
\end{eqnarray}
where the new spectral parameter $v_{k}$ is defined by:
\begin{eqnarray}
v_{k}:=y_{k}+\frac{1}{y_{k}}.
\end{eqnarray}
For completeness we state the final solutions of $M$ and $N$:
\begin{eqnarray}
M&=& \frac{\frac{i}{g}}{v_{1}-v_{2}-\frac{i}{g}}\nonumber\\
N&=& -\frac{v_{1}-v_{2}}{v_{1}-v_{2}-\frac{i}{g}}.
\end{eqnarray}
The last thing to determine is the function $h$. Since we have an
exact formula for $M$ and $N$, finding the solution is rather
straightforward.
\begin{eqnarray}
h(y_{1},y_{2})&=&-\frac{i}{y_{1}y_{2}}\frac{y_{1}-y_{2}}{v_{1}-v_{2}-\frac{i}{g}}.
\end{eqnarray}
Let us stress that the solutions obtained are unique.

In \cite{Beisert:2005tm} it is discussed how to generalize this to
more than two excitations by using the supersymmetry generators.
For this we need to consider the explicit four-dimensional
representation of $\mathfrak{su}(2|2)$ and in particular the
representation of the supersymmetry generators
$\mathcal{Q}_{\alpha}$. We use the conventions from
\cite{Arutyunov:2006yd}:
\begin{eqnarray}
\mathcal{Q}_{k,\alpha}|0\rangle^{\mathrm{II}}&=&a_{k}|A_{1}^{\dag}(p_{1})\ldots
A_{\alpha}^{\dag}(p_{k}) \ldots
A_{1}^{\dag}(p_{K^{\mathrm{I}}})\rangle^{\mathrm{I}}\nonumber\\
\mathcal{Q}_{k,\alpha}\mathcal{Q}_{l,\beta}|0\rangle^{\mathrm{II}}&=&a_{k}a_{l}|A_{1}^{\dag}(p_{1})\ldots
A_{\alpha}^{\dag}(p_{k}) \ldots A_{\beta}^{\dag}(p_{l}) \ldots
A_{1}^{\dag}(p_{K^{\mathrm{I}}})\rangle^{\mathrm{I}}\nonumber\\
\mathcal{Q}_{k,\alpha}\mathcal{Q}_{k,\beta}|0\rangle^{\mathrm{II}}&=&a_{k}b_{k}\epsilon^{\alpha\beta}|A_{1}^{\dag}(p_{1})\ldots
A_{2}^{\dag}(p_{k}) \ldots
A_{1}^{\dag}(p_{K^{\mathrm{I}}})\rangle^{\mathrm{I}},
\end{eqnarray}
with
$a_{k}=\sqrt{g}\sqrt{i(x_{k}^{-}-x_{k}^{+})}e^{i\frac{p_{k+1}+\ldots+
p_{K^{\mathrm{I}}}}{2}}$ and $b_{k}=-\frac{a_{k}}{x_{k}^{-}}$. We
define dressed supersymmetry generators:
\begin{eqnarray}
\mathcal{Q}_{\alpha,k}^{\pm}:=
e^{-i\frac{P}{2}}\frac{x^{\pm}_{k}}{x^{\pm}_{k}-x^{\mp}_{k}}\mathcal{Q}_{\alpha,k},
\end{eqnarray}
where $P:=\sum_{i=1}^{K^{\mathrm{I}}}p_{i}$. By using the
identity:
\begin{eqnarray}
\frac{y}{y-x^{-}_{k}}=\frac{x^{+}_{k}}{x^{+}_{k}-x^{-}_{k}}
+\frac{x^{-}_{k}}{x^{-}_{k}-x^{+}_{k}}
\frac{y-x_{k}^{+}}{y-x^{-}_{k}},
\end{eqnarray}
we see that we can write the one excitation state as:
\begin{eqnarray}
\sum_{k=0}^{K^{\mathrm{I}}}\Phi_{k}
(\mathcal{Q}^{-}_{\alpha,k}+\mathcal{Q}^{+}_{\alpha,k+1})|0\rangle^{\mathrm{II}},
\qquad \Phi_{k}:=\prod_{l=1}^{k}\frac{y-x_{l}^{+}}{y-x^{-}_{l}}.
\end{eqnarray}
This formula can be seen as a level II excitation which is moved
through the vacuum via $\Phi_{k}$, where it can be joined with the
vacuum to the left by $\mathcal{Q}^{-}$ or to the right by
$\mathcal{Q}^{+}$. The two-excitation state can now be written as
\begin{eqnarray}
|A^{\dag}_{\alpha}A^{\dag}_{\beta}\rangle^{\mathrm{II}}&=&
\frac{1}{2}\sum_{k=0}^{K^{\mathrm{I}}}\Phi_{k}(y_{1})\Phi_{k}(y_{1})\left\{\mathcal{Q}^{-}_{\alpha,k}
\mathcal{Q}^{-}_{\beta,k}+2\mathcal{Q}^{-}_{\alpha,k}
\mathcal{Q}^{+}_{\beta,k+1}\right.\nonumber\\
&&\qquad\left.+\mathcal{Q}^{+}_{\alpha,k+1}
\mathcal{Q}^{+}_{\beta,k+1}\right\}|0\rangle^{\mathrm{II}}\\
&&+\sum_{k<l=0}^{K^{\mathrm{I}}}\Phi_{k}(y_{1})\Phi_{l}(y_{2})
(\mathcal{Q}^{-}_{\alpha,k}+\mathcal{Q}^{+}_{\alpha,k+1})(\mathcal{Q}^{-}_{\beta,l}+\mathcal{Q}^{+}_{\beta,l+1})|0\rangle^{\mathrm{II}}\nonumber.
\end{eqnarray}
The first term is asymmetric since we need to make sure that
$y_{1}$ stays to the left of $y_{2}$, so the first term can be
seen as the ordered version of the second. The total
two-excitation state is now given by:
\begin{eqnarray}
|2\rangle^{\mathrm{II}}=
|A^{\dag}_{\alpha}A^{\dag}_{\beta}\rangle^{\mathrm{II}}+S^{\mathrm{II}}_{12}|A^{\dag}_{\alpha}A^{\dag}_{\beta}\rangle^{\mathrm{II}},
\end{eqnarray}
with
\begin{eqnarray}
S^{\mathrm{II}}_{12}|A^{\dag}_{\alpha}(y_{1})A^{\dag}_{\beta}(y_{2})\rangle^{\mathrm{II}}=
M|A^{\dag}_{\alpha}(y_{2})A^{\dag}_{\beta}(y_{1})\rangle^{\mathrm{II}}+NS^{\mathrm{II}}_{12}|A^{\dag}_{\beta}(y_{2})A^{\dag}_{\alpha}(y_{1})\rangle^{\mathrm{II}}.
\end{eqnarray}
From this one can completely get rid of the explicit use of
$A_{2}^{\dag}$ in the formulae, since the corresponding factor has
been distributed among the two different regions. This is now
easily generalized to an arbitrary number of excitations.

Finally, by now introducing the third level vacuum (\ref{VacIII}),
we can compute $S^{\mathrm{II,II}}$. It follows that:
\begin{eqnarray}
S^{\mathrm{II,II}}=-M-N=1.
\end{eqnarray}
Note that we follow the convention of \cite{Beisert:2005tm} and
introduce and additional $-$ sign when we scatter two fermions.

\subsection{Final Levels}

All that remains is a brief derivation of the last terms. The
procedure is exactly the same as above, only the expressions
involved are considerably more simple. We define the next
reference state to be:
\begin{eqnarray}
|0\rangle^{\mathrm{III}}=|A^{\dag}_{3}(y_{1})\ldots
A^{\dag}_{3}(y_{{K^{\mathrm{II}}}})\rangle^{\mathrm{II}}
\end{eqnarray}
and we only need to consider the creation operators $A^{\dag}_{4}$
as excitations (note that by the discussion at the end of the
previous section, we tacitly split up the $A_{2}^{\dag}$ into an
$A_{3}^{\dag}$ and an $A_{4}^{\dag}$). Repeating the process
described above, we are led to define a one-excitation state:
\begin{eqnarray}
|A^{\dag}_{4}(w)\rangle^{\mathrm{III}} =
\sum_{k=1}^{K^{\mathrm{II}}}\Psi^{(2)}_{k}(w)|A^{\dag}_{3}(y_{1})\ldots
A^{\dag}_{4}(y_{k})\ldots
A^{\dag}_{3}(y_{K^{\mathrm{II}}})\rangle^{\mathrm{II}},
\end{eqnarray}
with
\begin{eqnarray}
\Psi_{k}^{(2)}(w)=f^{(2)}(w,y_{k})\prod_{l=1}^{k-1}S^{\mathrm{III},\mathrm{II}}(w,y_{k}).
\end{eqnarray}
Note that, with a modest amount of foresight, we have already
included the explicit dependence on the pseudo-momentum $w$, which
will again come in as an integration constant. This time, the
scattering relations are given by (\ref{ScatteringII}) and the
compatibility relation is given by:
\begin{eqnarray}
S^{\mathrm{II}}_{(k,l)}|A^{\dag}_{4}(w)\rangle =
S^{\mathrm{II,II}}|A^{\dag}_{4}(w)\rangle_{(k,l)},
\end{eqnarray}
where this time the subscript $(k,l)$ stands for interchanging
$y_{k}$ and $y_{l}$. This yields the equations:
\begin{eqnarray}
M f^{(2)}(w,v_{1}) + N
f^{(2)}(w,v_{2})S^{\mathrm{III},\mathrm{II}}(w,y_{1}) &=&
f^{(2)}(w,v_{2})\nonumber\\
N f^{(2)}(w,v_{1}) + M
f^{(2)}(w,v_{2})S^{\mathrm{III},\mathrm{II}}(w,y_{1}) &=&
f^{(2)}(w,v_{1})S^{\mathrm{III},\mathrm{II}}(w,y_{2}),
\end{eqnarray}
in which the conventional extra $-$ is to be read in $M,N$ and
$S^{\mathrm{II,II}}$. These equations are straightforwardly solved
by:
\begin{eqnarray}
f^{(2)}(w,y_{k})&=&\frac{w-\frac{i}{2g}}{w-v_{k}-\frac{i}{2g}}\nonumber\\
S^{\mathrm{III},\mathrm{II}}(w,y_{k})&=&\frac{w-v_{k}+\frac{i}{2g}}{w-v_{k}-\frac{i}{2g}}.
\end{eqnarray}
When we consider a two excitation state, we will need to introduce
the S-matrix, $S^{\mathrm{III,III}}(w_{1},w_{2})$, which governs
the interchanging of the $w$s, analogous to the previous level.
The same ansatz for the two-excitation state as in the previous
level can be made (without the term in which the excitations are
on the same position, of course):
\begin{eqnarray}
|A_{4}^{\dag}(w_{1})A_{4}^{\dag}(w_{2})\rangle^{\mathrm{III}}=\sum_{l_{1}<l_{2}}\Psi^{(2)}_{l_{1}}(w_{1})\Psi^{(2)}_{l_{2}}(w_{2})
|\ldots A^{\dag}_{4}(y_{l_{1}})\ldots
A^{\dag}_{4}(y_{l_{2}})\ldots\rangle^{\mathrm{II}}.
\end{eqnarray}
By imposing the compatibility condition on the generic two
excitation state at this level
\begin{eqnarray}
|A^{\dag}_{4}(w_{1})A^{\dag}_{4}(w_{2})\rangle^{\mathrm{III}}+
S^{\mathrm{III,III}}
|A^{\dag}_{4}(w_{1})A^{\dag}_{4}(w_{2})\rangle^{\mathrm{III}}
\end{eqnarray}
one derives the following equation:
\begin{eqnarray}
f^{(2)}_{11}f^{(2)}_{22}S^{\mathrm{III,II}}_{21}+
S^{\mathrm{III,III}}f^{(2)}_{21}f^{(2)}_{12}S^{\mathrm{III,II}}_{12}=\nonumber\\
f^{(2)}_{12}f^{(2)}_{21}S^{\mathrm{III,II}}_{22}+
S^{\mathrm{III,III}}f^{(2)}_{22}f^{(2)}_{11}S^{\mathrm{III,II}}_{11}.
\end{eqnarray}
This is solved by:
\begin{eqnarray}
S^{\mathrm{III,III}}(w_{1},w_{2}) =
\frac{w_{1}-w_{2}-\frac{i}{g}}{w_{1}-w_{2}+\frac{i}{g}},
\end{eqnarray}
where again an additional $-$ sign was absorbed. In general the
$K^{\mathrm{III}}$-excitation state is given by:
\begin{eqnarray}\label{BALvl3}
|A_{4}^{\dag}(w_{1})\ldots
A_{4}^{\dag}(w_{K^{\mathrm{III}}})\rangle^{\mathrm{III}} +
S^{\mathrm{III}}\cdot |A_{4}^{\dag}(w_{1})\ldots
A_{4}^{\dag}(w_{K^{\mathrm{III}}})\rangle^{\mathrm{III}},
\end{eqnarray}
with
\begin{eqnarray}
&&|A_{4}^{\dag}(w_{1})\ldots
A_{4}^{\dag}(w_{K^{\mathrm{III}}})\rangle^{\mathrm{III}} =\\
&&\ \sum_{l_{1}<\ldots<
l_{K^{\mathrm{III}}}}\Psi^{(2)}_{l_{1}}(w_{1})\ldots\Psi^{(2)}_{l_{K^{\mathrm{III}}}}(w_{K^{\mathrm{III}}})
|A^{\dag}_{3}(y_{1})\ldots A^{\dag}_{4}(y_{l_{1}})\ldots
A^{\dag}_{4}(y_{l_{K^{\mathrm{III}}}})\ldots\rangle^{\mathrm{II}}\nonumber.
\end{eqnarray}
Now the creation operator picture has completely dissolved and we
are only left with the numbers
$K^{\mathrm{I}},K^{\mathrm{II}},K^{\mathrm{III}}$ and the momenta
$p,y,w$. From the above discussion, it is easily seen that
\begin{eqnarray}
K^{\mathrm{I}}&=&N(A^{\dag}_{1})+N(A^{\dag}_{2})+N(A^{\dag}_{3})+N(A^{\dag}_{4})\nonumber\\
K^{\mathrm{II}}&=&2N(A^{\dag}_{2})+N(A^{\dag}_{3})+N(A^{\dag}_{4})\nonumber\\
K^{\mathrm{III}}&=&N(A^{\dag}_{2})+N(A^{\dag}_{4}),
\end{eqnarray}
where $N(A^{\dag}_{M})$ stands for the number of $A^{\dag}_{M}$s
in the state. $K^{\mathrm{I}}$ is the number of creation
operators, $K^{\mathrm{II}}$ is the fermion number and
$K^{\mathrm{III}}$ is the number of fermions of flavor
$A_{4}^{\dag}$. One can use these numbers in the above ansatz for
the wave function and go back through all levels to obtain the
total wave function of the system corresponding to $p,y,w$. Thus,
one starts with a level III wave function with $K^{\mathrm{III}}$
excitations and by (\ref{BALvl3}) one writes this as a linear
combination of level II states, which can be written in terms of
level I states. This final result is, by construction, a solution
of (\ref{BAE}).

\subsection{Comparison with the Spin Chain Picture}

In \cite{Beisert:2005tm}, dynamic spin chains are considered. As
discussed before, the S-matrix, is given by (\ref{AFZSmatrix}).
When comparing the above discussion to the one in
\cite{Beisert:2005tm}, there are a few notational issues to have
in mind. First, we have $A^{\dag}_{a}\leftrightarrow\phi^{a}$ and
$A^{\dag}_{\alpha}\leftrightarrow\psi^{\alpha}$. Also, since in
this case one deals with spin chains, there are also $\mathcal{Z}$
fields present in the discussion. This alters the first Bethe
ansatz, in the sense that there is now a level \textrm{I} vacuum,
consisting only of $\mathcal{Z}$s.

Other than this, the entire discussion basically goes through,
apart from the fact that the spin chains are dynamic. This means
that $\mathcal{Z}$ fields can be created and annihilated by
creation operators $\mathcal{Z}^{+}$ and annihilation operators
$\mathcal{Z}^{-}$. These operators give an additional phase factor
in (\ref{TwoExcState}), which, in this picture, becomes:
\begin{eqnarray}
|\Psi^{\mathrm{II}}\rangle &=&
f(y_{1},x_{1})f(y_{2},x_{2})S^{\mathrm{II,I}}(y_{2},x_{1})|\psi^{\alpha}_{1}\psi^{\beta}_{2}\rangle^{\mathrm{I}}\nonumber\\
&&+f(y_{1},x_{1})f(y_{2},x_{1})f(y_{1},y_{2},x_{1})\frac{x_{2}^{-}}{x_{2}^{+}}\epsilon^{\alpha\beta}|\phi^{2}_{1} \phi^{1}_{1}\mathcal{Z}^{+}\rangle^{\mathrm{I}}\nonumber\\
&&+f(y_{1},x_{2})f(y_{2},x_{2})S^{\mathrm{II,I}}(y_{1},x_{1})S^{\mathrm{II,I}}(y_{2},x_{1})f(y_{1},y_{2},x_{2})\epsilon^{\alpha\beta}| \phi^{1}_{1}\phi^{2}_{1}\mathcal{Z}^{+}\rangle^{\mathrm{I}}\nonumber\\
&&+M(y_{1},y_{2})f(y_{2},x_{1})f(y_{1},x_{2})S^{\mathrm{II,I}}(y_{1},x_{1})|\psi^{\alpha}_{1}\psi^{\beta}_{2}\rangle^{\mathrm{I}}\nonumber\\
&&+N(y_{1},y_{2})f(y_{2},x_{1})f(y_{1},x_{2})S^{\mathrm{II,I}}(y_{1},x_{1})|\psi^{\beta}_{1}\psi^{\alpha}_{2}\rangle^{\mathrm{I}}.
\end{eqnarray}
The results for the spin chain are given by:
\begin{eqnarray}
\begin{array}{lll}
  S^{\mathrm{I,I}}=S_{0}(p_{k},p_{l})\frac{x_{l}^{-}-x_{k}^{+}}{x_{l}^{+}-x_{k}^{-}} & &S^{\mathrm{III,II}}=\frac{w-v_{k}+\frac{i}{2g}}{w-v_{k}-\frac{i}{2g}}\\
  S^{\mathrm{II,I}}=\frac{y-x_{k}^{+}}{y-x_{k}^{-}} &  & S^{\mathrm{III,III}}=\frac{w_{1}-w_{2}-\frac{i}{g}}{w_{1}-w_{2}+\frac{i}{g}} \\
  S^{\mathrm{II,II}}=1.&
  &
\end{array}
\end{eqnarray}
Since the string and spin chain S-matrix only differ by
$x$-dependent phase factors, one expects that the different levels
of the S-matrix also only differ by phase factors, which is indeed
the case. Furthermore, the factors depending only on the auxiliary
parameters $y,w$ coincide, which is not surprising since they are
independent of $x$. Finally, note that $\mathcal{Z}$ fields
basically give an extra level and hence one obtains in this case
an extra Bethe equation, which corresponds to the level-matching
condition (zero world-sheet momentum).

\section{Bethe Equations}

\subsection{Final Level}

\begin{figure}
\begin{center}
\psfrag{Tag1}[][l]{0} \psfrag{Tag2}[][l]{$L_{\mathrm{III}}$}
\psfrag{Tag3}[][l]{$2L_{\mathrm{III}}$}
\psfrag{Tag4}[][l]{$3L_{\mathrm{III}}$}
\psfrag{Tag5}[][l]{$w_{1}$} \psfrag{Tag6}[][l]{$w_{2}$}
\psfrag{Tag7}[][l]{$w_{3}$} \psfrag{Tag8}[][l]{$w_{4}$}
\includegraphics{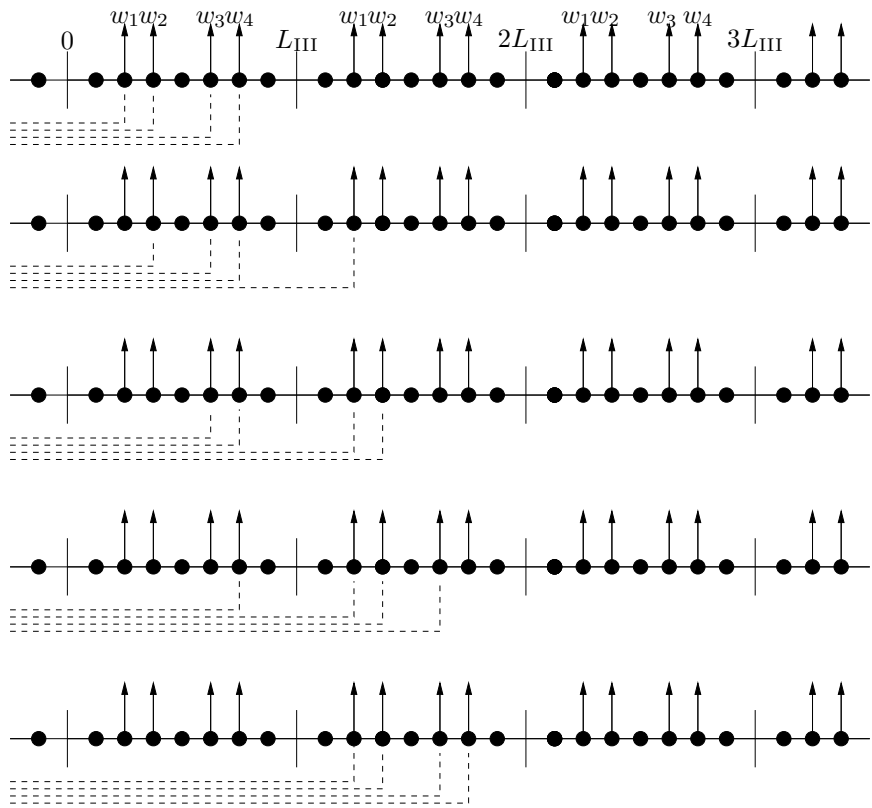}
\end{center}
\caption{Schematic representation of the periodicity. The plain
dots represent the operators forming the vacuum and the arrows
stand for the excitations.}\label{FigBAE}
\end{figure}

In this section we will derive the Bethe equations, by imposing
periodicity on the wave function. Consider a chain with
$K^{\mathrm{II}}$ sectors and $K^{\mathrm{III}}$ excitations. The
way to impose periodicity is depicted in Figure \ref{FigBAE}.

In this figure the dots represent the $K^{\mathrm{II}}$ sites and
the arrows represent the excitations at this level. The drawn
configuration depicts one term of the ansatz for the wave
function:
\begin{eqnarray}
|A_{4}^{\dag}(w_{1})\ldots
A_{4}^{\dag}(w_{K^{\mathrm{III}}})\rangle^{\mathrm{III}} +
S^{\mathrm{III}}|A_{4}^{\dag}(w_{1})\ldots
A_{4}^{\dag}(w_{K^{\mathrm{III}}})\rangle^{\mathrm{III}},
\end{eqnarray}
where $S^{\mathrm{III}}|A_{4}^{\dag}(w_{1})\ldots
A_{4}^{\dag}(w_{K^{\mathrm{III}}})\rangle^{\mathrm{III}}$ stands
for the different configurations corresponding to interchanging
the $w$s. Furthermore, we have:
\begin{eqnarray}
&&|A_{4}^{\dag}(w_{1})\ldots
A_{4}^{\dag}(w_{K^{\mathrm{III}}})\rangle^{\mathrm{III}}=\\
\quad
&&\sum_{l_{1}<\ldots<l_{K^{\mathrm{III}}}}\Psi^{(2)}_{l_{1}}(w_{1})\ldots\Psi^{(2)}_{l_{K^{\mathrm{III}}}}(w_{K^{\mathrm{III}}})
|A_{4}^{\dag}(y_{1})\ldots A_{4}^{\dag}(y_{l_{1}})\ldots
A_{3}^{\dag}(y_{K^{\mathrm{II}}})\rangle^{\mathrm{II}}\nonumber.
\end{eqnarray}
Recall that the coefficients $\Psi^{(2)}_{l}$ are given by
\begin{eqnarray}
\Psi^{(2)}_{l}(w)=f^{(2)}(v_{l},w)\prod_{l}S^{\mathrm{III,II}}(v_{i},w),
\end{eqnarray}
but since we have periodicity, there is an ambiguity in choosing
over which sites the product is taken. This is represented by the
dotted lines in Figure \ref{FigBAE}. The different depicted
choices lead to a consistency check, corresponding to periodicity.
Note that this is just the analogue of (\ref{period}).

For each of the choices one can write down the explicit wave
function, by just following the prescription given in the previous
sections. We will give the explicit formulas for the first two
lines, leaving the other ones for the interested reader.

The first configuration is just the superposition of
$K^{\mathrm{III}}$ plane waves:
\begin{eqnarray}
\sum_{l_{1}<\ldots<l_{K^{\mathrm{III}}}}\Psi^{(2)}_{l_{1}}(w_{1})\ldots\Psi^{(2)}_{l_{K^{\mathrm{III}}}}(w_{K^{\mathrm{III}}})
|A_{4}^{\dag}(y_{1})\ldots A_{4}^{\dag}(y_{l_{1}})\ldots
A_{3}^{\dag}(y_{K^{\mathrm{II}}})\rangle^{\mathrm{II}}.
\end{eqnarray}
The second wave function is also a superposition of plane waves,
but this time, the parameters $w$ are in different order and we
pick up additional factors of $S^{\mathrm{III,II}}$:
\begin{eqnarray}
&&S^{\mathrm{III,II}}(w_{1},v_{1})\ldots
S^{\mathrm{III,II}}(w_{1},v_{K^{\mathrm{II}}})\prod_{l\neq1}^{K^{\mathrm{III}}}S^{\mathrm{III,III}}(w_{1},w_{l})\times\\
\quad&&\times\sum_{l_{1}<\ldots<l_{K^{\mathrm{III}}}}\Psi^{(2)}_{l_{1}}(w_{2})\ldots\Psi^{(2)}_{l_{K^{\mathrm{III}}}}(w_{1})
|A_{4}^{\dag}(y_{1})\ldots A_{4}^{\dag}(y_{l_{1}})\ldots
A_{3}^{\dag}(y_{K^{\mathrm{II}}})\rangle^{\mathrm{II}}\nonumber.
\end{eqnarray}
The factors of $S^{\mathrm{III,III}}(w_{l},w_{1})$ arise because
$w_{1}$ is now to the right of the other $w$s. Since these two
wave functions should be equal, it is easy to see that the
following equation should hold:
\begin{eqnarray}
\prod_{l=1}^{K^{\mathrm{II}}}S^{\mathrm{II,III}}(v_{l},w_{1})\prod_{l\neq1}^{K^{\mathrm{III}}}S^{\mathrm{III,III}}(w_{l},w_{1})=1,
\end{eqnarray}
where we define
$S^{\mathrm{II,III}}(v_{l},w_{k}):=\frac{1}{S^{\mathrm{III,II}}(w_{k},v_{l})}$.
By considering the other choices one easily derives all the
$K^{\mathrm{III}}$ Bethe equations for this level:
\begin{eqnarray}
\prod_{l=1}^{K^{\mathrm{II}}}S^{\mathrm{II,III}}(v_{l},w_{k})\prod_{l\neq
k }^{K^{\mathrm{III}}}S^{\mathrm{III,III}}(w_{l},w_{k})=1.
\end{eqnarray}
We see that this coincides with (\ref{RealBAE}). Finally, note
that from these equations it also follows that the choice of
origin is irrelevant as is seen by comparing the first and last
line in Figure \ref{FigBAE}, in which all the dotted lines are in
the interval $[L,2L]$ opposed to the interval $[0,L]$.

\subsection{Other Levels}

\begin{figure}
\begin{center}
\psfrag{Tag1}[][l]{0} \psfrag{Tag2}[][l]{$L$}
\psfrag{Tag3}[][l]{$2L$} \psfrag{Tag4}[][l]{$3L$}
\psfrag{Tag5}[][l]{} \psfrag{Tag6}[][l]{$y_{1}$}
\psfrag{Tag7}[][l]{$y_{1}$} \psfrag{Tag8}[][l]{$\binom{y_{3}}{w}$}
\includegraphics{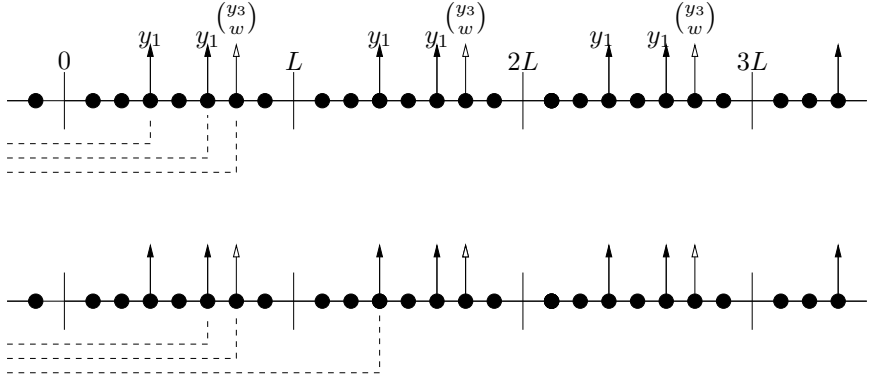}
\end{center}
\caption{Schematic representation of the periodicity at the second
level. The plain dots represent the operators forming the vacuum
at this level. The black arrows stand for the $A_{3}$ excitations
and the white arrow corresponds to the $A_{4}$
excitation.}\label{FigBAE2}
\end{figure}

In this section we will only treat the second level Bethe
equations. The level I Bethe equations are, of course, obtained in
a similar way. We apply the same procedure as above. The only
difference is that we have more types of excitations. For ease of
survey, we will only consider an explicit example, leaving the
general case, which is not much more difficult, for the interested
reader. We consider the case with two $A_{3}^{\dag}$ and one
$A_{4}^{\dag}$ operator, see Figure \ref{FigBAE2}.

Again, the shown configurations in Figure \ref{FigBAE2} only
correspond to one term from the full wave function. Let us write
the full wave function to make things more explicit. We have
$K^{\mathrm{III}}=1$, so our wave function at this level is:
\begin{eqnarray}
|A_{4}^{\dag}\rangle^{\mathrm{III}}.
\end{eqnarray}
We can write this out in a linear combination of level II terms,
as explained above:
\begin{eqnarray}
|A_{4}^{\dag}\rangle^{\mathrm{III}}=
\Psi^{(2)}_{1}(w)|A_{4}^{\dag}A_{3}^{\dag}A_{3}^{\dag}\rangle^{\mathrm{II}}+
\Psi^{(2)}_{2}(w)|A_{3}^{\dag}A_{4}^{\dag}A_{3}^{\dag}\rangle^{\mathrm{II}}+
\Psi^{(2)}_{3}(w)|A_{3}^{\dag}A_{3}^{\dag}A_{4}^{\dag}\rangle^{\mathrm{II}}.
\end{eqnarray}
Then we can do the same for each level II term and we obtain a
large linear combination of level I wave functions,
\begin{eqnarray}
|A_{4}^{\dag}A_{3}^{\dag}A_{3}^{\dag}\rangle^{\mathrm{II}}=
\sum_{l_{1}<l_{2}<l_{3}}\Psi_{l_{1}}(y_{1})\Psi_{l_{2}}(y_{2})\Psi_{l_{3}}(y_{3})|\ldots
A_{4}^{\dag}\ldots A_{3}^{\dag}\ldots
A_{3}^{\dag}\ldots\rangle^{\mathrm{I}} +\ldots
\end{eqnarray}
One such a term from this expansion is depicted in Figure
\ref{FigBAE2}.

From the previous section, we know that the level III wave function
can be written down unambiguously if the Bethe equations at that
level are satisfied. However, for the level II wave function these
equations still need to be derived. The procedure is completely
analogous to the one given above for $\Psi^{(2)}$ and the result
is given by
\begin{eqnarray}
\prod_{m=1}^{K^{\mathrm{I}}}S^{\mathrm{II,I}}(y_{k},x_{m})\prod_{l\neq
k}^{K^{\mathrm{II}}}S^{\mathrm{II,II}}(y_{k},y_{l})
\prod_{n=1}^{K^{\mathrm{III}}}S^{\mathrm{II,III}}(w_{k},y_{n})=1,
\end{eqnarray}
which coincides with (\ref{RealBAE}). Note that in the derivation
of this equation, compatibility of the level III state under the
level II S-matrix plays a crucial role. Finally, when comparing to
(\ref{RealBAE}), we use the definition:
\begin{eqnarray}
S^{AB}(x_{k}^{A},x_{l}^{B})=\frac{1}{S^{BA}(x_{l}^{B},x_{k}^{A})}.
\end{eqnarray}
The Bethe equation can be read as follows. We take the second line
and we permute the level II excitation back to its original
position. Doing this, we have to permute it past a level III
excitation, giving a factor $S^{\mathrm{II,III}}(w_{l},y_{k})$,
get it past level II excitations, which give
$S^{\mathrm{II,II}}(y_{l},y_{k})$ and finally we have moved it
through the vacuum, giving the $S^{\mathrm{II,I}}(y_{k},x_{l})$
terms.

\subsection{Results}

Let us derive the explicit Bethe equations. We will first give
them for the $\mathfrak{su}(2|2)$ case, from which the
$\mathfrak{su}(2|2)^{2}$ follows.

From (\ref{RealBAE}) we obtain the following three Bethe
equations:
\begin{eqnarray}
e^{ip_{k}(L+\frac{K^{\mathrm{I}}}{2}-\frac{K^{\mathrm{II}}}{2})}&=&
e^{i\frac{P}{2}}\prod_{l=1,l\neq k
}^{K^{\mathrm{I}}}\left[S_{0}(p_{k},p_{l})\frac{x_{k}^{+}-x_{l}^{-}}{x_{k}^{-}-x_{l}^{+}}\right]
\prod_{l=1}^{K^{\mathrm{II}}}\frac{{x_{k}^{-}-y_{l}}}{x_{k}^{+}-y_{l}} \nonumber\\
1&=&e^{-i\frac{P}{2}}\prod_{l=1}^{K^{\mathrm{I}}}\frac{y_{k}-x^{+}_{l}}{y_{k}-x^{-}_{l}}
\prod_{l=1}^{K^{\mathrm{III}}}\frac{y_{k}+\frac{1}{y_{k}}-w_{l}+\frac{i}{2g}}{y_{k}+\frac{1}{y_{k}}-w_{l}-\frac{i}{2g}}\nonumber\\
1&=&\prod_{l=1}^{K^{\mathrm{II}}}\frac{w_{k}-y_{l}-\frac{1}{y_{l}}+\frac{i}{2g}}{w_{k}-y_{l}-\frac{1}{y_{l}}-\frac{i}{2g}}
\prod_{l=1,l\neq
k}^{K^{\mathrm{III}}}\frac{w_{k}-w_{l}-\frac{i}{g}}{w_{k}-w_{l}+\frac{i}{g}}.
\end{eqnarray}
These equations coincide, as it should be, with the ones given
\cite{Martins:2007hb}. The equations for the full
$\mathfrak{su}(2|2)^{2}$ case follow easily:
\begin{eqnarray}
e^{ip_{k}(L+K^{\mathrm{I}}-\frac{K_{(1)}^{\mathrm{II}}}{2}-\frac{K_{(2)}^{\mathrm{II}}}{2})}&=&
e^{iP}\prod_{l=1,l\neq k
}^{K^{\mathrm{I}}}\left[S_{0}(p_{k},p_{l})\frac{x_{k}^{+}-x_{l}^{-}}{x_{k}^{-}-x_{l}^{+}}\right]^{2}\times\nonumber\\
&&\times\prod_{\alpha=1}^{2}\prod_{l=1}^{K_{(\alpha)}^{\mathrm{II}}}\frac{{x_{k}^{-}-y^{(\alpha)}_{l}}}{x_{k}^{+}-y^{(\alpha)}_{l}} \\
1&=&e^{-i\frac{P}{2}}\prod_{l=1}^{K^{\mathrm{I}}}\frac{y^{(\alpha)}_{k}-x^{+}_{l}}{y^{(\alpha)}_{k}-x^{-}_{l}}
\prod_{l=1}^{K_{(\alpha)}^{\mathrm{III}}}\frac{y_{k}^{(\alpha)}+\frac{1}{y_{k}^{(\alpha)}}-w_{l}^{(\alpha)}+\frac{i}{2g}}{y_{k}^{(\alpha)}+\frac{1}{y_{k}^{(\alpha)}}-w_{l}^{(\alpha)}-\frac{i}{2g}}\nonumber\\
1&=&\prod_{l=1}^{K_{(\alpha)}^{\mathrm{II}}}\frac{w_{k}^{(\alpha)}-y_{k}^{(\alpha)}-\frac{1}{y_{k}^{(\alpha)}}+\frac{i}{2g}}{w_{k}^{(\alpha)}-y_{k}^{(\alpha)}-\frac{1}{y_{k}^{(\alpha)}}-\frac{i}{2g}}
\prod_{l\neq
k}^{K_{(\alpha)}^{\mathrm{III}}}\frac{w_{k}^{(\alpha)}-w_{l}^{(\alpha)}-\frac{i}{g}}{w_{k}^{(\alpha)}-w_{l}^{\alpha}+\frac{i}{g}}\nonumber,
\end{eqnarray}
with $\alpha=1,2$.

Let us conclude by comparing these equations to the ones proposed
in \cite{Beisert:2005fw}. We compare our equations to the
reformulated version of these equations in
\cite{Hentschel:2007xn}. We see that they agree with the
$(\eta_{1},\eta_{2})=(+,+)$ sector if one imposes the level
matching condition $e^{iP}=1$ and if one makes the following
identifications:
\begin{eqnarray}
(K^{\mathrm{I}},K^{\mathrm{II}}_{(1)},K^{\mathrm{II}}_{(2)},K^{\mathrm{III}}_{(1)},K^{\mathrm{III}}_{(2)})&=&
(K_{4},K_{1}+K_{3},K_{5}+K_{7},K_{2},K_{6})\\
(x_{k}^{\pm};y^{(1)}_{k};y^{(2)}_{k};v^{(1)}_{k};v^{(2)}_{k};w^{(1)}_{k};w^{(2)}_{k})&=&
(x_{4,k}^{\pm};x_{3,k};x_{5,k};u_{3,k};u_{5,k};u_{2,k};u_{6,k})\nonumber,
\end{eqnarray}
with the parameter $L=J$.

\section{Conclusions}

In this note, the Bethe equations corresponding to the string
S-matrix from \cite{Arutyunov:2006yd} were derived by using the
coordinate Bethe ansatz. The equations obtained for the string
case correspond to the ones recently obtained in
\cite{Martins:2007hb} and also coincide with the proposed
equations in \cite{Beisert:2005fw}. The method was already applied
to the spin chain case in \cite{Beisert:2005tm}.

It would be interesting to study the relation between both
approaches to the Bethe ansatz, especially the relation between
the FZ creation operators and the creation operators obtained from
the monodromy matrix in the algebraic Bethe ansatz.

There still remains a lot to be studied about these equations. The
dependence on the momenta and total momentum $P$ may yield
interesting results, for example in the thermodynamic limit.
Furthermore, it would be interesting to see if the equations can
be obtained with the method of \cite{Kazakov:2007fy}.

Finally, since it is possible from the S-matrix from
\cite{Arutyunov:2006yd} to link the spin chain picture with the
string picture, it would be nice to study if both cases can be
linked via a continuous family of S-matrices and if the
aforementioned procedure can be applied.

\section*{Acknowledgements}

We are grateful to G. Arutyunov, B. Eden, S. Frolov and J. Leow,
for valuable discussions. This work was supported in part by the
EU-RTN network \textit{Constituents, Fundamental Forces and
Symmetries of the Universe} (MRTN-CT-2004-005104), by the INTAS
contract 03-51-6346 and by the NWO grant 047017015.

\appendix

\section{S-matrix}

For completeness, let us give the explicit form of coefficients of
the S-matrix (\ref{AFZSmatrix}):
\begin{eqnarray}
a_{1}(p_{1},p_{2})&=&S_{0}(p_{1},p_{2})\frac{x_{2}^{-}-x_{1}^{+}}{x_{2}^{+}-x_{1}^{-}}\frac{\eta_{1}\eta_{2}}{\widetilde{\eta}_{1}\widetilde{\eta}_{2}}\nonumber\\
a_{2}(p_{1},p_{2})&=&S_{0}(p_{1},p_{2})\frac{(x_{1}^{-}-x_{1}^{+})(x_{2}^{-}-x_{2}^{+})(x_{2}^{-}+x_{1}^{+})}{(x_{1}^{-}-x_{2}^{+})(x_{1}^{-}x_{2}^{-}-x_{1}^{+}x_{2}^{+})}\frac{\eta_{1}\eta_{2}}{\widetilde{\eta}_{1}\widetilde{\eta}_{2}}\nonumber\\
a_{3}(p_{1},p_{2})&=&-S_{0}(p_{1},p_{2})\nonumber\\
a_{4}(p_{1},p_{2})&=&S_{0}(p_{1},p_{2})\frac{(x_{1}^{-}-x_{1}^{+})(x_{2}^{-}-x_{2}^{+})(x_{1}^{-}+x_{2}^{+})}{(x_{1}^{-}-x_{2}^{+})(x_{1}^{-}x_{2}^{-}-x_{1}^{+}x_{2}^{+})}\nonumber\\
a_{5}(p_{1},p_{2})&=&S_{0}(p_{1},p_{2})\frac{x_{2}^{-}-x_{1}^{-}}{x_{2}^{+}-x_{1}^{-}}\frac{\eta_{1}}{\widetilde{\eta}_{1}}\nonumber\\
a_{6}(p_{1},p_{2})&=&S_{0}(p_{1},p_{2})\frac{x_{1}^{+}-x_{2}^{+}}{x_{1}^{-}-x_{2}^{+}}\frac{\eta_{2}}{\widetilde{\eta}_{2}}\nonumber\\
a_{7}(p_{1},p_{2})&=&iS_{0}(p_{1},p_{2})\frac{(x_{1}^{-}-x_{1}^{+})(x_{2}^{-}-x_{2}^{+})(x_{1}^{+}-x_{2}^{+})}{(x_{1}^{-}-x_{2}^{+})(1-x_{1}^{-}x_{2}^{-})\widetilde{\eta}_{1}\widetilde{\eta}_{2}}\nonumber\\
a_{8}(p_{1},p_{2})&=&iS_{0}(p_{1},p_{2})\frac{x_{1}^{-}x_{2}^{-}(x_{1}^{+}-x_{2}^{+})\eta_{1}\eta_{2}}{x_{1}^{+}x_{2}^{+}(x_{1}^{-}-x_{2}^{+})(1-x_{1}^{-}x_{2}^{-})}\nonumber\\
a_{9}(p_{1},p_{2})&=&S_{0}(p_{1},p_{2})\frac{x_{1}^{+}-x_{1}^{-}}{x_{1}^{-}-x_{2}^{+}}\frac{\eta_{2}}{\widetilde{\eta}_{1}}\nonumber\\
a_{10}(p_{1},p_{2})&=&S_{0}(p_{1},p_{2})\frac{x_{2}^{+}-x_{2}^{-}}{x_{1}^{-}-x_{2}^{+}}\frac{\eta_{1}}{\widetilde{\eta}_{2}}.
\end{eqnarray}
The parameters $x^{\pm}_{k}$ are related to the quasi-momentum of
the magnons and the coupling constant in the standard way:
\begin{eqnarray}\label{parameters}
\frac{x^{+}_{k}}{x^{-}_{k}}=e^{ip_{k}},\quad
x^{+}_{k}+\frac{1}{x^{+}_{k}}-x^{-}_{k}-\frac{1}{x^{-}_{k}}=\frac{i}{g}.
\end{eqnarray}
Furthermore, the scalar function $S_{0}(p_{1},p_{2})$ is of the
form:
\begin{eqnarray}
S_{0}(p_{1},p_{2})^{2}=\frac{x_{2}^{+}-x_{1}^{-}}{x_{2}^{-}-x_{1}^{+}}\frac{1-\frac{1}{x_{1}^{+}x_{2}^{-}}}{1-\frac{1}{x_{1}^{-}x_{2}^{+}}}e^{i\theta(p_{1},p_{2})}e^{ia(p_{1}\epsilon_{2}-p_{2}\epsilon_{1})}.
\end{eqnarray}
This S-matrix encompasses both the spin chain S-matrix and the
string S-matrix, depending on the choice of parameters. To be
precise, the spin chain S-matrix is given by making the following
choice:
\begin{eqnarray}
\eta_{1}&=&\eta(p_{1})\nonumber\\
\eta_{2}&=&\eta(p_{2})\nonumber\\
\widetilde{\eta}_{1}&=&\eta(p_{1})\nonumber\\
\widetilde{\eta}_{2}&=&\eta(p_{2}).
\end{eqnarray}
The string S-matrix is obtained by choosing:
\begin{eqnarray}
\eta_{1}&=&\eta(p_{1})e^{i\frac{p_{2}}{2}}\nonumber\\
\eta_{2}&=&\eta(p_{2})\nonumber\\
\widetilde{\eta}_{1}&=&\eta(p_{1})\nonumber\\
\widetilde{\eta}_{2}&=&\eta(p_{2})e^{i\frac{p_{1}}{2}}.
\end{eqnarray}
In both cases, $\eta(p_{k}):=\sqrt{i(x^{-}_{k}-x^{+}_{k})}$.

\bibliographystyle{h-physrev}
\bibliography{LiteratureSpin}

\end{document}